\documentclass[review,12pt]{elsarticle}

\usepackage{amssymb}
\usepackage{amsmath}
\usepackage{amsthm,amsmath}
\usepackage{natbib}
\RequirePackage{hyperref}
\hypersetup{
    colorlinks=true, 
    linkcolor=black, 
    citecolor=black, 
    filecolor=black, 
    urlcolor=black   
}
\usepackage[utf8]{inputenc} 
\usepackage{graphicx}
\usepackage{booktabs}
\usepackage{xcolor}
\usepackage{csquotes} 
\usepackage{url}
\usepackage{rotating}

\journal{Online Social Networks and Media}

\begin{document}

\begin{frontmatter}



\title{Influencer Self-Disclosure Practices on Instagram: A Multi-Country Longitudinal Study}

\author[uu]{Thales Bertaglia}
\author[uu]{Catalina Goanta}
\author[um]{Gerasimos Spanakis}
\author[um]{Adriana~Iamnitchi}

\affiliation[uu]{organization={Utrecht University},
            city={Utrecht},
            country={the Netherlands}}

\affiliation[um]{organization={Maastricht University},
            city={Maastricht},
            country={the Netherlands}}
\begin{abstract}
This paper presents a longitudinal study of more than ten years of activity on Instagram consisting of over a million posts by 400 content creators from four countries: the US, Brazil, Netherlands and Germany. 
Our study shows differences in the professionalisation of content monetisation between countries, yet consistent patterns; significant differences in the frequency of posts yet similar user engagement trends; and significant differences in the disclosure of sponsored content in some countries, with a direct connection with national legislation. We analyse shifts in marketing strategies due to legislative and platform feature changes, focusing on how content creators adapt disclosure methods to different legal environments. We also analyse the impact of disclosures and sponsored posts on engagement and conclude that, although sponsored posts have lower engagement on average, properly disclosing ads does not reduce engagement further. Our observations stress the importance of disclosure compliance and can guide authorities in developing and monitoring them more effectively.
\end{abstract}




\end{frontmatter}



\section{Introduction}

The social media economy is changing dramatically. 
In their early history (the mid-2000s), social networks were predominantly used for peer-to-peer communication and user-generated content that reflected people's daily lives~\cite{boyd2007}. 
At that point, business models and monetisation opportunities were still largely undeveloped. 
The advertising potential of harnessing human attention in social surroundings saw the rise of individual Internet entrepreneurs who have since received the label of social media influencers. In this study, we refer to influencers as social media content creators who monetise online content through native advertising~\cite{EUP-influencers, Wojdynski2015}. 

To deepen the understanding of the evolving ecosystem of content monetisation on social media, we conducted a longitudinal study spanning 2010-2020 of 400 Instagram influencers from four distinct countries. We present an analysis of the professionalisation of their advertising activities, focusing on the evolution of influencer self-disclosure practices. For this purpose, we collected and analysed a dataset of more than one million posts by 50 mega and 50 micro-influencers each from Brazil, Germany, the Netherlands, and the US. These countries were selected based on their different regulatory approaches in terms of legal standards and enforcement, which we will discuss in~\autoref{sec:regulations}.   

Our contributions include (1)~a methodological approach to longitudinal data curation in the context of influencer selection, laying the foundations for more reproducible and robust research in the field of influencer studies; (2)~empirical observations of the rise and decline in the volume of posts by these influencers over the years and the effect on disclosed ads; (3)~a comparative analysis of different disclosure practices for Instagram posts, likely correlated with national regulation; (4)~a characterisation of monetisation strategies in terms of the timing of posts and partnership with global brands; (5)~a thorough analysis of the impact of disclosures on engagement; and (6)~an investigation of the prevalence of undisclosed ads in our dataset and their effect on engagement. Our study provides a foundation for updating regulatory guidelines and developing automated tools to enhance transparency and compliance in influencer marketing by offering insights into how disclosure practices and content strategies impact audience engagement. Ultimately, these observations can inform legislation and public policy on strategies that may lead to changes in practices related to ad disclosures on social media.

\section{Background}

Disclosures are a crucial legal aspect of influencer marketing. This section presents a brief overview of the general legal framework for advertising in the four countries in our dataset and discusses work related to our research.

\subsection{Legal Background on Advertising}
\label{sec:regulations}
The regulation of social media advertising by influencers is more straightforward than public debates may lead us to believe. 
For decades, technology-driven advertising developments (e.g. radio, television) have posed different iterations of the same legal question: is it fair to hide advertising? 
The answer is a strong no. 
As early as 1937, the US Federal Trade Commission (FTC) recognised the harms of deceptive advertising (FTC v. Standard Education Society, 1937). 
Consumers may generally be sceptical of advertising but not of information perceived as independent or unrelated to advertising~\cite{petty2013-puffery,advertising-theory2019}.

In their quest to develop new business models, social media platforms have incorporated new ways to embed hidden advertising into social media content. This poses several risks, including manipulating consumers through misleading information and hurting commercial competitors who do not engage in misleading practices.  
Legally speaking, deceptive practices pose issues at the advertising intersection of consumer and media law, often leading to the so-called \enquote{mandatory rules}. These are rules that all market actors must respect and cannot contract against. 

However, in addition to laws adopted by the state, the advertising industry has also promoted and developed extensive self-regulation.
For instance, in the US, in 2006, the FTC adopted its Endorsement Guides~\cite{FTC-guides} applicable to social media, which were updated in 2022 to include more specific references to influencer marketing; in the Netherlands, the Stichting Reclame Code adopted a Social Media Advertising Code, updated in 2022 to the Social Media and Influencer Marketing Advertising Code~\cite{SRC}. 
Similarly, Brazil adopted its Digital Influencer Advertising Guidelines in 2021~\cite{conar-br}.

While codes of conduct and guidelines feature detailed standards (e.g., what hashtags to use for disclosure and influencer definitions), they often do not have a mandatory status. 
Consequently, enforcing these rules in practice is difficult, as additional administrative and judicial actions are necessary. 

Germany stands out in terms of disclosure regulation due to the high volume of cases brought before German courts about non-disclosed advertising. 
Only on 9 September 2021, the German Supreme Court (\textit{Bundesgerichtshof}) came out with three judgments on influencer marketing~\cite{BGH}.  
Among the clarifications addressed in these cases, the Court discussed the fact that advertising disclosures should not be included in a \enquote{hashtag cloud} that would reduce their clarity. 
In addition, the Court indicated the need to use unambiguous disclosure terms such as \enquote{Anzeige} or \enquote{Werbung}, as for German-speaking consumers, foreign language terms such as \enquote{advertisement} or \enquote{collaboration} would not be considered sufficient. 
These rulings, often focused on Instagram, clarify a divergent landscape of lower court decisions, sometimes going so far as to say that all posts should be tagged as ads due to the commercial nature of influencer accounts~\cite{LG_Berlin_2018}. 
This initial confusion regarding disclosure compliance nurtured pragmatic disclosure approaches.    

Interestingly, the laws on unfair practices that apply in Germany are also applicable in the Netherlands due to European harmonisation~\cite{directive_ec}. 
However, Dutch courts have not yet had the opportunity to analyse unfair practices concerning influencer marketing. 
The socio-legal reasons behind these differences are too complex to map or speculate upon here. 
Factors such as compliance perception, trust in the market, maturity of the market, contractual setup, business diligence, access to legal knowledge, and care for consumers can all explain why some influencers choose to disclose advertising more than others. 

From a regulatory perspective, Germany, the Netherlands, the US, and Brazil offer a wide enough range of disclosure self-regulation and mandatory rules applicable on three continents. While German regulations have been notably strict, it is essential to note that regulations in other countries have evolved differently. The US, for instance, saw early adoption of endorsement guidelines by the FTC, but the enforcement intensity has varied over time. The Netherlands has developed a comprehensive self-regulation framework, although it lacks extensive judicial analysis. Brazil's guidelines are relatively recent, adopted in 2021, and it remains to be seen how they will be enforced in practice. In addition, the United States is often seen as the ground zero of social media policy development, not only because the most prominent social media platforms are based in California, but also because of the very early popularity of influencer marketing in the United States. 

\subsection{Related Work on Influencer Studies}
Influencer marketing has been the initial monetisation model used by influencers. It involves influencers providing advertising services for various commercial (and often non-commercial) actors. 
So far, studies in the field have focused on identifying influencers, describing their characteristics, and mapping the prevalence of their disclosures. Although influencer studies and research agendas have gained attention in a wide range of disciplines, e.g. communication, law, media studies, and computer science,
they are generally focused on specific issues, communities and data. Qualitative studies are predominant within the social sciences and 
have primarily focused on questions such as the effects of content monetisation for brands, how influencers perceive and react to metrics and disclosure requirements, and on analysing the relationship between influencers and their followers~\cite{belancheUnderstandingInfluencerMarketing2021a,mishraConsumerDecisionmakingOmnichannel2021,tanwarTrendsInfluencerMarketing2022,TrustMeTrust,vrontisSocialMediaInfluencer2021, arriagadaYouNeedLeast2020, christinDramaMetricsStatus2021,blut2023effectiveness,hudders2021commercialization}, among others. 
On the other hand, large-scale data-driven studies similar to ours are still rare within the field.

Most notably, Mathur et al.~\cite{mathurEndorsementsSocialMedia2018} analyse disclosures in over 2.6 million publications on YouTube and Pinterest and find that only 10\% of content sponsored through affiliate marketing has any form of disclosures. Although their work focuses on understanding the prevalence and efficacy of disclosures from a regulatory perspective, their analysis focuses on English-language content from the US and does not include other jurisdictions. 

In the context of Instagram, Yang et al.~\cite{yang2020mention} focus on the dynamics of brand mentions in sponsored posts. They analyse a network of over 18,000 influencers and their brand mentions on Instagram. Their study highlights that most influencers typically mention just a few select brands. While high-profile influencers gravitate towards well-known brands, micro-influencers do not exhibit a particular preference. Additionally, the study observes that audience engagement remains remarkably similar between sponsored and non-sponsored posts. Martins et al.~\cite{martins2022characterizing} focus on characterising the evolution of influencer marketing on Facebook and Instagram, especially in the context of the COVID-19 pandemic. Their study analyses 9.5 million sponsored posts and reveals shifting trends in ads and user engagement after the pandemic's onset across various topics, including sports, retail, and politics. 

Many empirical studies focus on applying machine learning methods to detect sponsored posts and characterise influencers. Zarei et al.~\cite{zareiCharacterisingDetectingSponsored2020} compile a dataset of 35 thousand posts and 99 thousand stories by 12 thousand different users for training machine learning models to detect undisclosed ads; all posts were selected based on containing hashtags related to sponsored content (e.g. \#ad). They characterise the activity of influencers of different sizes (\textit{nano}, \textit{micro}, and \textit{mega}) through cumulative distribution functions of different features, such as the number of followers, likes, and the type of product being advertised. Enforcing transparency in influencer marketing has posed significant challenges despite increasing efforts by global regulatory bodies. A particular area of focus is the automated detection of sponsored content. Bertaglia et al.~\cite{bertaglia2023closing} discuss the issues surrounding this task, especially the inconsistencies arising from human data annotation for machine learning models. 

Similarly, Kim et al.~\cite{kimDiscoveringUndisclosedPaid2021b} introduce a dataset with 1.6 million posts by 38 thousand influencers and use different network features, including brand mentions as connections between posts, to train deep learning models to detect undisclosed ads. Kim et al.~\cite{kimMultimodalPostAttentive2020} use a version of the same dataset to profile influencers according to the industries related to the products they advertise. While these studies employ large-scale datasets, their data curation methodologies differ from ours, primarily focusing on ad identification rather than longitudinal characterisation. This approach, which often selects posts and accounts based on specific hashtags, can make it more challenging to discern the representation of languages, countries, and account types in the dataset. Moreover, given their focus, these datasets typically cover shorter time frames, which may not be tailored for extensive longitudinal analyses like ours.

Our study fills that gap by focusing on a curated list of content creators restricted to selected countries and collecting the entire post history for all accounts. While previous work investigated what is \textit{not} being disclosed (by automatically detecting such content), our work focuses on understanding influencer self-disclosure practices and their connection to regulation. 

The few previous studies focusing on regulation \cite{ershovEffectsInfluencerAdvertising2020} analysed the relation between disclosed and undisclosed ads and regulations from Germany and Spain. They collect posts from 12 thousand influencers (6 thousand from each country) and compare the volume of sponsored content and engagement before and after regulatory changes. They find reductions in engagement and activity, measured as the number of sponsored posts. However, this study diverges from our primary emphasis on influencer self-disclosure practices, and its data curation methodology is not explicitly detailed. In addition, it only focuses on two European countries.

Similarly, Waltenrath~\cite{waltenrathEmpiricalEvidenceImpact2021a} looks at disclosures and engagement penalties and shows that advertising posts attract only about 75\% of the engagement of regular posts. Our study aims to develop these insights further by investigating how content creators express disclosures using different strategies and how that relates to regulations from different countries.

\section{Dataset}
\label{sec:dataset}

Because our objective is a longitudinal study of monetisation practices and disclosures, we collected our dataset using an influencer-focused approach. 
Publicly available Instagram datasets (such as the one by Kim et al.~\cite{kimDiscoveringUndisclosedPaid2021b}) were collected for different purposes using a hashtag-focused approach and do not provide adequate temporal data for a longitudinal study; moreover, they are generally not curated for particular countries.

We first selected four countries with different regulatory frameworks to allow for a diverse sample of self-disclosure advertising practices: Brazil (BR), Germany (DE), the Netherlands (NL), and the United States (US).  
We used Heepsy\footnote{\url{heepsy.com}}, one of the many platforms that have emerged to support influencer marketing, to select influencers from each of these countries based on the number of followers. 
We chose Heepsy because it offers a location-based selection filter.

We focused on two groups of influencers: those with more than $500K$ followers, which we call \emph{mega-influencers}, and those with fewer than $500K$ followers, which we call \emph{micro-influencers}. The choice of the $500k$ cutoff was inspired by a 2022 national oversight measure implemented by the Dutch Media Authority that focuses on video platform accounts with more than $500k$ followers\footnote{{https://www.cvdm.nl/nieuws/commissariaat-voor-de-media-start-toezicht-op-video-uploaders/}}. 
While various other cutoffs exist in the literature, there is no standard definition of what a mega or micro-influencer is. Moreover, follower volume remains a random measurement for influencer studies, as the importance of size varies between platforms and industries. As monetisation models evolve and influencers diversify their strategies, traditional definitions around influencer categories, especially regarding follower count, become increasingly blurred--particularly considering the recent rise of nano and niche influencers~\cite{wibawa2021role, lie2022effect}. We chose the 500k follower cutoff to align with our study’s objective of understanding how influencers’ visibility impacts their adherence to regulations. Influencers with more than 500k followers are highly visible and likely face more scrutiny, which might drive them to adhere more strictly to advertising regulations. While influencers around the 500k mark are also significantly visible, we hypothesise that those surpassing this threshold experience a qualitative change in their professional practices and engagement with brands, thus attracting more regulatory attention. This hypothesis is supported by the focus of the Dutch Media Authority on accounts above this threshold, suggesting that such influencers are considered impactful enough to warrant specific oversight. Therefore, the 500k cutoff provides a meaningful distinction to study differences in compliance and professionalisation.

To ensure a comprehensive analysis, we initially aimed to select 400 accounts evenly distributed across the four countries and between mega- and micro- categories. Having established our criteria, we used Heepsy to search for accounts that matched our specifications. Heepsy returns, by default, a list of accounts sorted \textit{relevance}; the details of this metric -- how it is defined, calculated and applied in the ranking process -- are unclear, as Heepsy offers no transparency regarding its algorithms. Bishop~\cite{bishop2021infmgttools} highlights that influencer management tools, such as Heepsy, often incorporate concepts related to \textit{brand risk} in their metrics, and these concepts could propagate and amplify biases related to sexuality, class, and race. By encoding subjective and potentially discriminatory criteria into their algorithms, platforms like Heepsy may privilege certain types of content over others, leading to a skewed and potentially biased representation of influencers. Therefore, to avoid the potential impact of such biases, we chose not to rely on Heepsy's proprietary sorting algorithm and instead sorted the accounts in decreasing order of the number of followers. Starting from the largest accounts, we went down the list in descending order of followers. This method caused our data to skew towards micro-influencers approaching the 500k cutoff, as we began selecting from the top of the descending list at 500k.

Heepsy also uses proprietary algorithms to infer the location of influencers. During the data selection process, we noticed that inferred countries were sometimes unreliable. 
Therefore, we manually curated each account and selected only those with explicit mentions of location and consistency in the language of the posts; this is not a widespread practice given the work-intensive curation and annotation task. Previous studies focus on identifying influencer accounts at scale based on hashtag selection~\cite{kimMultimodalPostAttentive2020,kimDiscoveringUndisclosedPaid2021b,zareiCharacterisingDetectingSponsored2020}. However, this approach may lead to inaccuracies that are difficult to map computationally. Although our approach may lead to the exclusion of outlier influencers not correctly identified by Heepsy, such a limitation exists across various fixed sampling methodologies. Systematically estimating the coverage of the data returned by Heepsy would require knowledge of every influencer based in a given country, which is not feasible given the current data access restrictions. However, our observations indicated that Heepsy correctly identified at least the most popular influencers from each country. 

Given the methodological challenges and the inherent biases of the Heepsy platform, we acknowledge the propensity of our sampling to skew towards micro-influencers nearing our $500k$ cutoff and mega-influencers with substantially larger followings. Obtaining a truly random sample of Instagram influencers is impossible, as we cannot access a complete list of users on the platform. To compile the final list of influencers, we excluded apparent accounts of organisations, fictional characters, pets, and \enquote{repost accounts} (profiles focused on compiling content from other users or platforms, e.g., memes).

We collected all posts and their metadata from each curated account using CrowdTangle, along with account-level information such as follower counts. A complete description of the information provided by CrowdTangle is available online\footnote{\url{help.crowdtangle.com/en/articles/4201940-about-us}}. 
We collected a total of $1,006,253$ posts spanning from October 2010 to September 2022. Of these, mega influencers contributed $568,915$ posts (56.5\%). Breaking it down by country, Brazilian influencers posted the most with $462,746$ posts (46.0\%), followed by the US with $294,599$ posts (29.3\%), the Netherlands with $143,978$ posts (14.3\%), and Germany with $104,930$ posts (10.4\%). \autoref{tab:dataset-stats} presents descriptive statistics on the most relevant features of the dataset. Each column represents a country (\textit{BR, DE, NL, US}) and size (\textit{Micro ($\mu$), Mega (M)}) combination (we call that a setting); \autoref{tab:agg-dataset-stats} presents the statistics combining data from all countries. 

\begin{table}[htbp]
	\caption{Dataset statistics. Values are averaged across all posts or accounts of the corresponding setting.}
\begin{tabular}{@{}lcccccccc@{}}
	\toprule
	& \multicolumn{2}{c}{\textbf{BR}} & \multicolumn{2}{c}{\textbf{DE}} & \multicolumn{2}{c}{\textbf{NL}} & \multicolumn{2}{c}{\textbf{US}} \\ \midrule
	& \textbf{$\mu$}  & \textbf{M} & \textbf{$\mu$}  & \textbf{M} & \textbf{$\mu$}  & \textbf{M} & \textbf{$\mu$}  & \textbf{M} \\
	\textbf{Followers} & 492k            & 12,552k        & 546k            & 5,767k        & 499k            & 2,041k        & 508k            & 26,641k \\
	\textbf{Verified}  & 56\%            & 100\%         & 62\%            & 96\%          & 66\%            & 90\%          & 66\%            & 100\% \\ \midrule
	\textbf{Posts}     & 4.7k             & 4.5k           & 1.0k             & 1.1k           & 1.0k             & 1.87k           & 2.0k             & 3.9k \\
	\textbf{Likes}     & 3.4k            & 69.6k        & 11.3k           & 101.5k        & 11.3k           & 25.8k         & 12.0k            & 178.7k \\
	\textbf{Comments}  & 65             & 1.5k          & 253             & 2.8k          & 155             & 271           & 155             & 3.0k \\ \bottomrule
\end{tabular}
\label{tab:dataset-stats}
\end{table}

\begin{table*}[htbp]
	\caption{Statistics of the dataset combining data from all countries. Values are averaged across all posts and accounts of the corresponding setting.} 
	\centering
\begin{tabular}{@{}lccc@{}}
	\toprule
	& \multicolumn{3}{c}{\textbf{Average}}                \\ \midrule
	  & \textbf{$\mu$} & \textbf{M} & \textbf{All} \\
	\textbf{Followers} & 511k           & 11,750k        & 6,130k         \\
	\textbf{Verified} & 63\%           & 97\%          & 80\%           \\ \midrule
	\textbf{Posts} & 2.2k            & 2.9k           & 2.5k            \\
	\textbf{Likes} & 5.7k           & 104.2k        & 60.8k          \\
	\textbf{Comments} & 110            & 1.9k          & 1.1k           \\ \bottomrule
\end{tabular}
\label{tab:agg-dataset-stats}
\end{table*}

There was a time lapse between our initial data curation (when we selected the accounts) and the final data collection stage. This interval caused some fluctuation in the number of followers. Consequently, a few micro-influencers surpassed the 500k followers mark when we compiled the data. However, these fluctuations were generally minor, averaging a 2\% deviation, and explain the average number of followers that exceeds the cutoff value in the case of DE and US in \autoref{tab:dataset-stats}. 

The average number of followers for micro-influencers is similar for all countries, with values close to $500k$, which matches the criterion used to select the influencers. 
The differences for mega-influencers are more significant and, at least partially, correlate to the countries' population size. 
Influencers from the US are more likely to have larger international audiences due to language and media influence, resulting in almost four times as many followers as the global average.
The \textit{Verified} row in~\autoref{tab:dataset-stats} denotes the percentage of accounts that have a verified badge, indicating that Instagram has confirmed the account's authenticity. 
Most mega-influencers have verified accounts, which is expected given their high number of followers.  

The total number of average posts per account varies greatly across settings and does not follow a clear pattern. Brazilian influencers (especially micro) are outliers, having a much higher post volume than the combined average. Similar differences have been identified on other platforms and are linked to the importance and pervasiveness of social media in Brazil and the Global South in general~\cite{rossiniDigitalMediaLandscape2021,seibel2021impact}. 
The average number of \textit{Likes} and \textit{Comments} per post, a measure of user engagement, correlates to the number of followers---a larger audience entails more people seeing and likely engaging with a post. It also correlates negatively to the number of posts, as the low relative engagement for Brazilian micro-influencers indicates. A high post-frequency might discourage followers from engaging because there is less novelty, or followers feel overloaded. 


To investigate the role of regulation in influencer marketing, we focus on disclosed posts because they directly impact how influencers frame their sponsored content. To identify self-disclosures, we compiled an initial list of \textit{hashtags} and \textit{keywords} frequently used to disclose ads based on previous work ~\cite{zareiCharacterisingDetectingSponsored2020,kimDiscoveringUndisclosedPaid2021b}. We expanded the original list to include related terms in other languages by analysing the co-occurrence of hashtags in a data sample. We also include a few non-standard spelling variations (e.g. typos). The final list contains \#ad, \#advertisement, \#advertising, \#advertisiment, \#advertisment, \#anzeige, \#publi, \#publicidade, \#publipost, \#werbung, \#spons,  \#samenwerking, and \#sponsored and their versions as keywords (without the \# symbol). 
These disclosure terms are typically included in the self-regulatory guidelines discussed in~\autoref{sec:regulations}. 
Finally, we labelled posts containing any of those terms as \textit{sponsored}. We also consider posts tagged with the \textit{paid partnership} label as sponsored. This label is an Instagram feature released in 2017 to encourage disclosing branded content.

\autoref{tab:sponsored-stats} presents statistics about sponsored posts aggregated by setting; \autoref{tab:agg-sponsored-stats} presents the statistics combining data from all countries.

\begin{table}[htbp]
	\caption{Sponsored posts statistics. All values are averaged across all posts/accounts of the corresponding setting. Sponsored accounts are considered those that have at least one sponsored post. Days First Sponsored represents the number of days between the very first post of the account and the first disclosed sponsored post.}
\begin{tabular}{lcccccccc}
	\hline
	& \multicolumn{2}{c}{\textbf{BR}} & \multicolumn{2}{c}{\textbf{DE}} & \multicolumn{2}{c}{\textbf{NL}} & \multicolumn{2}{c|}{\textbf{US}} \\ \hline
	& \textbf{$\mu$}  & \textbf{M} & \textbf{$\mu$}  & \textbf{M} & \textbf{$\mu$}  & \textbf{M} & \textbf{$\mu$}  & \textbf{M} \\
	\textbf{\# Spons. Posts}    & 2.9k             & 7.4k          & 9.3k            & 7.9k          & 730             & 989           & 1.4k             & 4.2k \\
	\textbf{\% Spons. Posts}    & 1.22            & 3.25          & 18.16            & 14.70          & 1.44            & 1.06          & 1.41            & 2.14 \\
	\textbf{Days First Spons.}   & 1603             & 1440           & 907             & 1022          & 1398            & 1527          & 1217             & 1134 \\
	\textbf{\% Spons. Accs.} & 88              & 98            & 88              & 78            & 84              & 74            & 86              & 96 \\ \hline
\end{tabular}
\label{tab:sponsored-stats}
\end{table}

\begin{table}[htbp]
	\caption{Sponsored posts statistics combining data from all countries. All values are averaged across all posts/accounts of the corresponding setting. Sponsored accounts are considered those that have at least one sponsored post. Days First Sponsored represents the number of days between the very first post of the account and the first disclosed sponsored post.}
	\centering
\begin{tabular}{lccc}
	\hline
	& \multicolumn{3}{c}{\textbf{Aggregated}} \\ \hline
	& \textbf{$\mu$} & \textbf{M} & \textbf{All} \\
	\textbf{\# Spons. Posts} & 14.3k           & 20.5k          & 34.8k \\
	\textbf{\% Spons. Posts}    & 3.27           & 3.60          & 3.45 \\
	\textbf{Days First Spons.}   & 1280            & 1279           & 1280 \\
	\textbf{\% Spons. Accs.} & 87             & 87            & 87 \\ \hline
\end{tabular}
\label{tab:agg-sponsored-stats}
\end{table}

We see that the number of sponsored posts varies greatly across countries. 
Germany comprises approximately 50\% of the total, indicating that their strict enforcement of regulation must affect disclosures. 
Brazilian micro-influencers have a significantly lower amount of disclosed sponsored posts. 
Combined with the high number of posts, this shows a lower level of professionalisation of monetisation practices; these accounts might not be getting sponsorships despite their frequent activity, or they might not be disclosing them, which connects to the lack of clear regulation and enforcement. 
The numbers from the Netherlands are also below the global average despite these accounts having a high relative engagement, which is often the measure brands use to find and employ influencers. 

\textit{Days First Sponsored} in \autoref{tab:sponsored-stats} represents the average number of days between the first post of an account and its first sponsored post.  This number shows how long influencers take to monetise their content and thus can serve as a proxy for professionalisation. 
Finally, we define an account as \textit{Sponsored} if it has at least one sponsored post, i.e., it shows whether the influencer has ever accepted paid sponsorships. 
The vast majority of influencers in all settings have sponsored posts. 
The number is consistent across sizes, indicating that micro-influencers appeal to brands as much as mega-influencers.


\section{Longitudinal Analysis}
\label{sec-longitudinal}

\begin{figure}[htbp]
        \centering
	\includegraphics[scale=0.55]{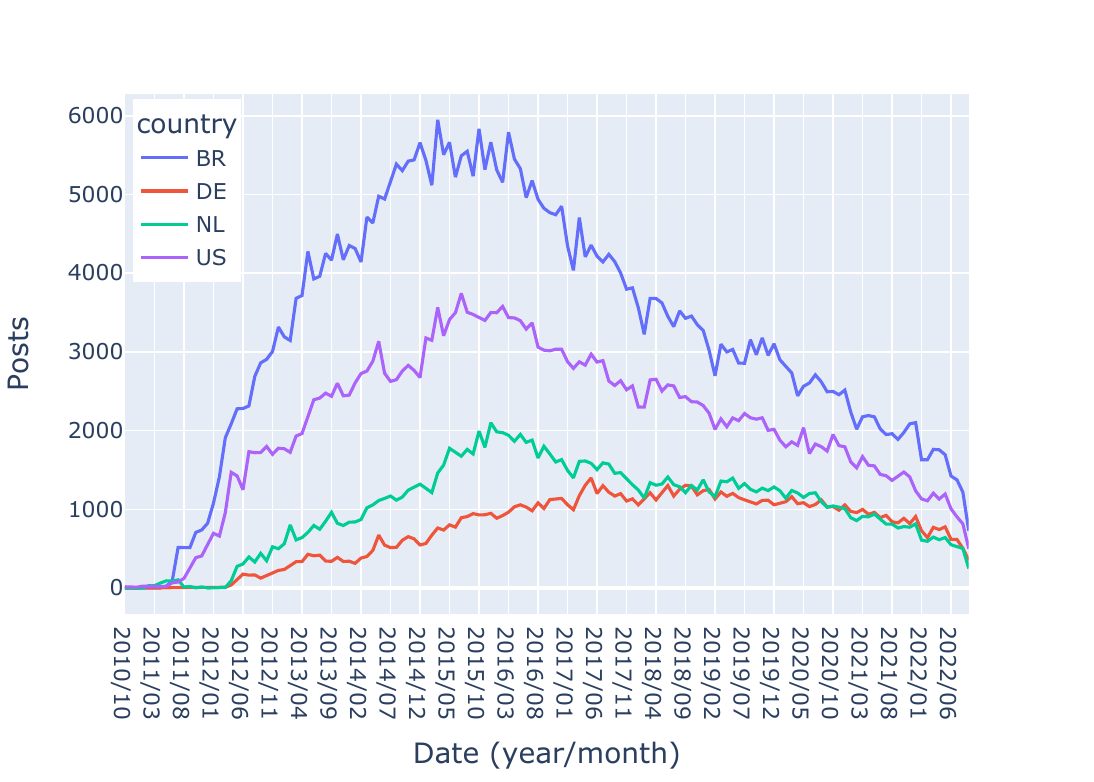}
	\caption{{Number of posts over time per country.}}
	\label{fig:posts-per-country}
\end{figure}

In order to investigate how content monetisation evolved on Instagram and to analyse trends over time, we conduct longitudinal analyses of different features of posts in our dataset. \autoref{fig:posts-per-country} presents the number of monthly posts per country throughout the observation period. 
Instagram was launched at the beginning of October 2010. 

The first posts in our dataset were from the US in October 2010, Brazil in December 2010, the Netherlands in February 2011, and Germany in September 2011. By mid-2011, the activity of the 100 accounts from Brazil significantly exceeded that of all other countries, with the US taking the second position.  
This ordering by the number of posts is maintained throughout the period and is also visible in Table~\ref{tab:dataset-stats}, where the average Brazilian account in our dataset makes approximately 4.5K posts, 50\% more than the average US account, and three times more than the average Dutch or German influencer account.  

With the exception of Germany, all countries saw a peak of activity in late 2015 and a consistent decrease in the monthly number of posts since then. The sharpest decline, which began around mid-2016, coincides with the release of the Stories feature in August 2016. Since our dataset only contains posts, the decrease in overall activity could indicate a shift in posting strategy -- from regular (permanent) posts to stories. 
The Brazilian accounts shrunk their activity fivefold in August 2022 from the peak in March 2015, while the German accounts only less than threefold. This difference highlights that although some activity might have shifted from posts to stories, the decline could at least partially be explained by platform migration -- some influencers might have increased their activity on other social media platforms, such as TikTok. In the context of Brazilian influencers, this trend contrasts with recent findings by Martins et al.~\cite{martins2022characterizing}, which indicate a consistent increase in the number of disclosed ads. We acknowledge this discrepancy and consider that factors such as differences in data collection and sampling methodologies may account for these differing observations. Unlike our approach, Martins et al.~\cite{martins2022characterizing} use a larger set of disclosure hashtags to categorise sponsored posts. Additionally, their criteria for selecting influencers differ from ours; they include accounts with over 1k followers that have posted disclosed content. Given our stricter definition, comparisons should be made with these differences in mind.
We verified that the accounts in our dataset were still active at the end of our observation period; however, we could not investigate their activity on other platforms. 

\begin{figure}[htbp]
        \centering
	\includegraphics[scale=0.55]{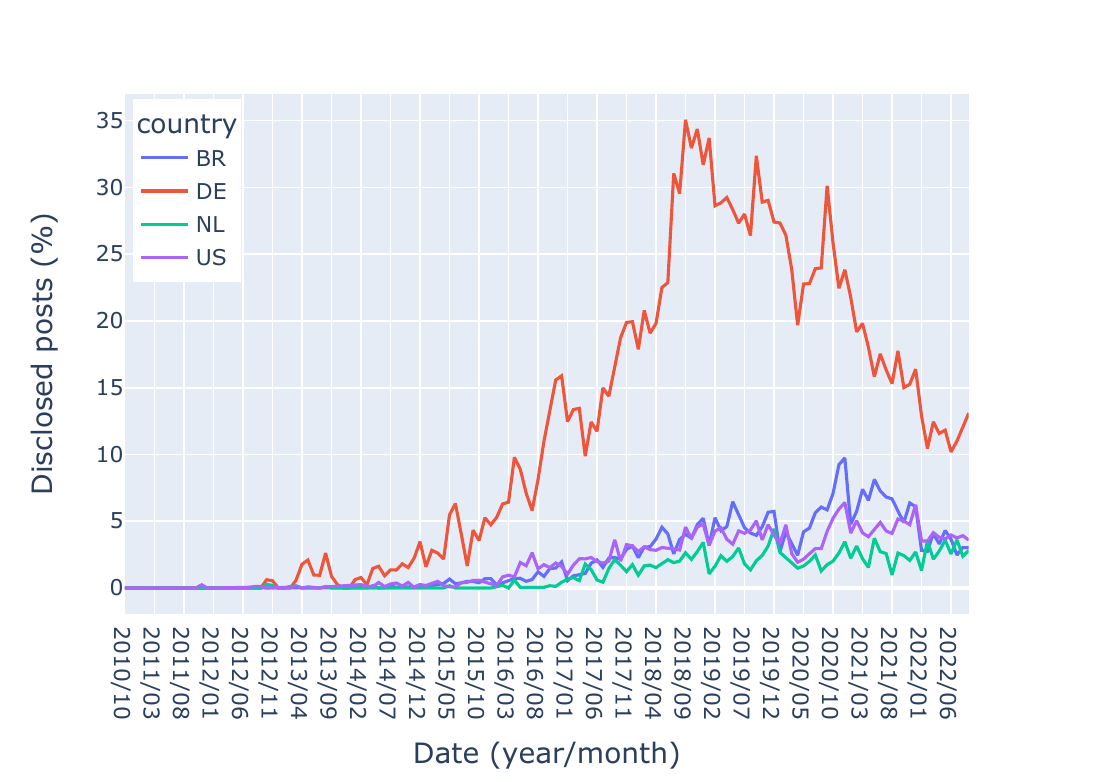}
	\caption{{Percentage of disclosed posts over time per country.}}
	\label{fig:disclosed-per-country}
\end{figure}

This pattern of activity is not noticeable in the time series of the percentage of disclosed sponsored posts shown in~\autoref{fig:disclosed-per-country}. 
As early as October 2012, two years after the launch of Instagram, disclosed sponsored posts were visible in the German influencers' activity, and they reached 35\% of all posts by October 2018. 
None of the other countries comes close to this level of disclosure. 

Although posts by German influencers comprise the majority of disclosures, the country has the least diversity in the number of accounts contributing to the total volume of disclosures. 17\% of accounts from Germany contribute with 80\% of the disclosures -- the lowest among all countries. This proportion shows that a few influencers concentrate most of the sponsored content, or at least are the ones to adhere more strictly to the disclosure requirements. Similarly, the other countries in the dataset, especially Brazil, show a higher diversity in the number of influencers with sponsored posts, which might indicate that the monetisation market is more accessible. 

\section{Ad Disclosure Strategies}
\label{sec:marketing-strategies}

Content creators use different strategies to disclose sponsored posts. Our dataset includes three types of disclosures: hashtags, keywords, and the paid partnership tag (AD tag). Different disclosure types impact users differently: Mathur et al.~\cite{mathurEndorsementsSocialMedia2018} show that the efficacy of disclosures on YouTube and Pinterest varies according to their type, and social media users particularly fail to understand nonexplanatory disclosures. We study how content creators interpret influencer marketing regulations and express their adherence to them through disclosures. 
We also investigate how different disclosure types impact user engagement. 

\begin{figure}[htbp]
        \centering
	\includegraphics[scale=.25]{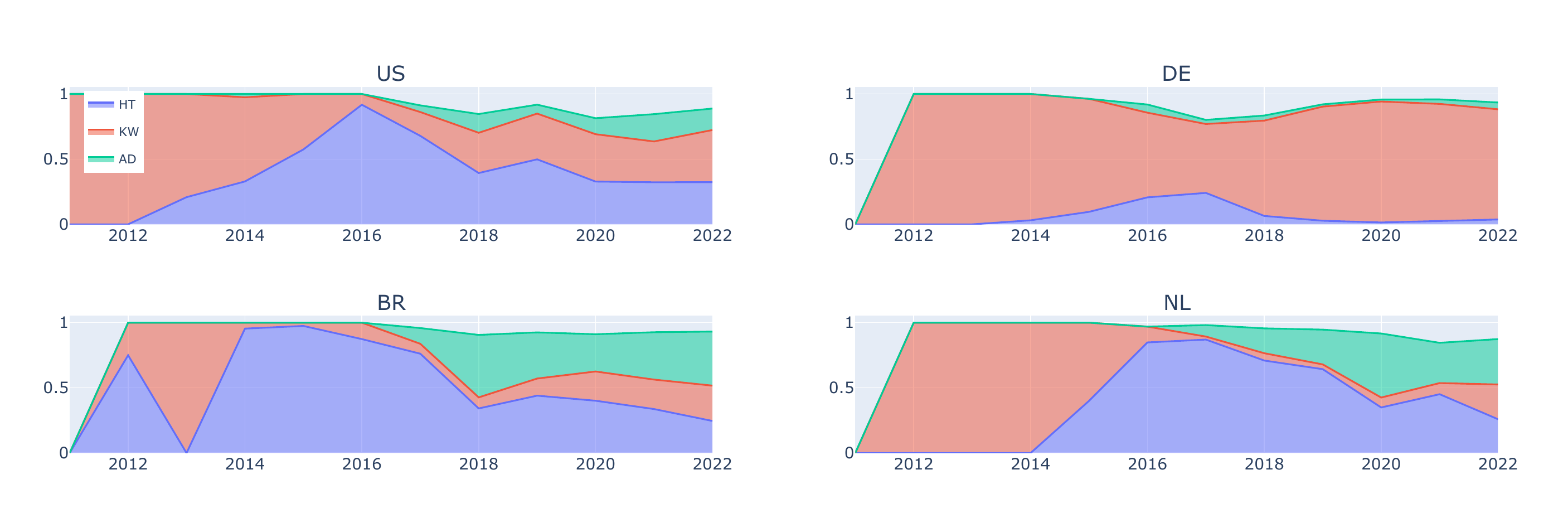}
		\caption{{Fraction of features used to disclose sponsored posts across countries over time. 
                About 80\% of all disclosed posts from DE use keywords (KW), while the majority of posts from the other countries use hashtags (HT). Paid partnership feature (AD) is used for 31\% of post in BR, 27\% in NL, 10\% in US and only 3\% in DE. Posts that use more than one disclosure method are not represented. There were no disclosed posts in 2010 and only US influencers disclosed sponsored posts in 2011}.}
			\label{fig:disclosure-type-time}
\end{figure}

To analyse how the use of disclosure types evolved over time, \autoref{fig:disclosure-type-time} shows the frequency of the three individual disclosure types over the years.

Starting in 2017, the use of the AD tag grew rapidly among BR and NL content creators. Instagram states that content creators must use the paid partnership label to tag the brand with which they are working. However, US and DE posts still mostly use hashtags and keywords to disclose sponsored content. From a user-experience perspective, the paid partnership label is the most visible disclosure type: posts with the tag have a clear label above the photo or video indicating that they are ads. Hashtags are primarily included at the final of a caption, after the entire text. Instagram's interface does not display long captions completely, so users must click to read the entire text and thus see the disclosure hashtags. Keywords could be placed anywhere within the caption. 

\begin{figure}[htbp]
        \centering
	\includegraphics[scale=.55]{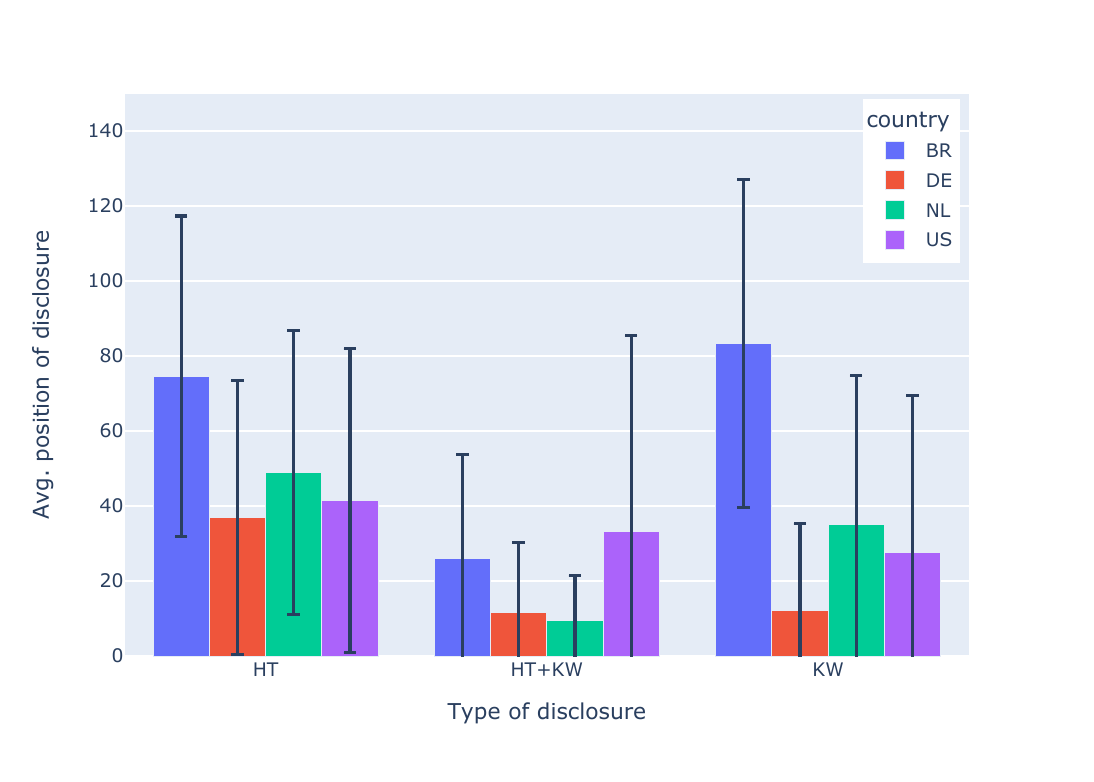}
		\caption{{Position of the first disclosure term in the caption, measured as the number of words from the beginning.}}
			\label{fig:disclosure-position-country}
\end{figure}

\autoref{fig:disclosure-position-country} presents the average position of the first disclosure in the captions as a proxy for the visibility of the disclosure types. The results are aggregated by country, with vertical lines indicating the standard deviation value.

On average, German content creators place disclosures at the start of the caption, highlighting a difference in positioning between keyword and hashtag disclosures.

\begin{figure}[htbp]
        \centering
	\includegraphics[scale=.3]{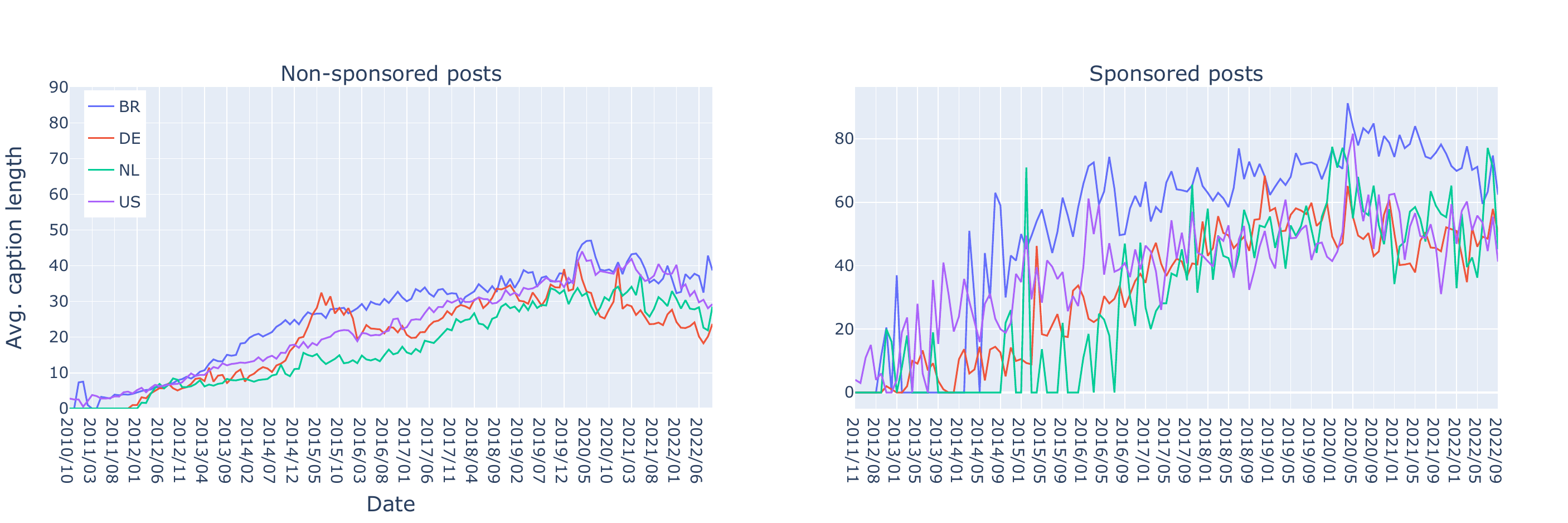}
		\caption{{Average number of words in a caption over time per country, per sponsored and non-sponsored posts.}}
			\label{fig:caption-len}
\end{figure}

Caption length also affects disclosure position, especially for hashtags -- longer captions increase the position of hashtags at the end.  \autoref{fig:caption-len} presents the average caption length per country for sponsored and non-sponsored posts.

The trend for both sponsored and non-disclosed posts shows that the average caption length increased significantly over the years in all countries. 
Sponsored posts tend to be much longer than non-disclosed ones, with nearly double the average length. 
This finding is consistent with the literature on marketing strategies, highlighting that advertising posts tend to include more complex narratives to persuade followers to engage with partner brands~\cite{lee2020influencer}. Despite the variations, the average disclosure position remained low, indicating that disclosures in posts by German content creators are more effective and consistent.

In general, our analyses show the influence of regulation on how content creators disclose paid sponsorship. 
With a prevalence of keyword disclosures, sponsored posts by German influencers tend to have more visible disclosures at the start of the captions.
Although the paid partnership label has gained traction in most countries in our dataset, it still falls behind other, less transparent, types of disclosure; one possible explanation for its lack of adoption is in its name: the wording implies a partnership involving only monetary transactions, failing to include other business models which are common in the context of influencer marketing, such as barter~\cite{goanta2020regulation}. 
Analysing content creators' use of disclosures helps evaluate regulation effectiveness and suggests ways to enhance disclosure visibility, ultimately improving transparency for content monetisation. 

\begin{figure}[htbp]
\centering
	\includegraphics[scale=0.55]{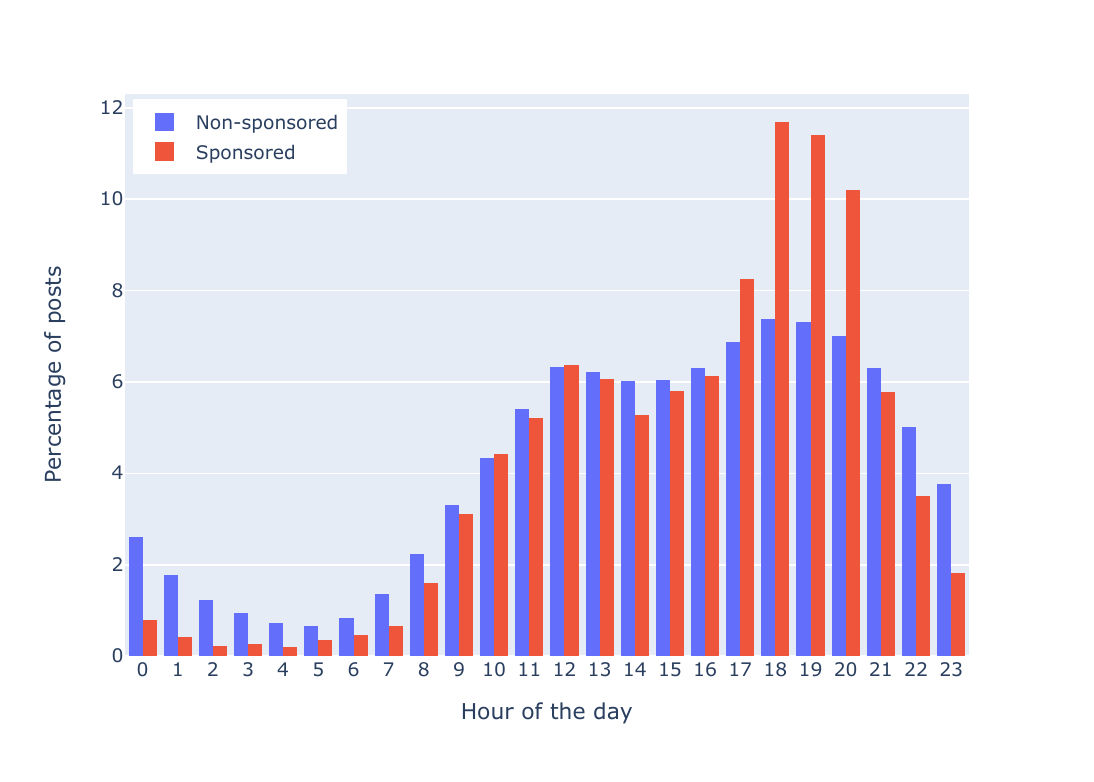}
	\caption{{Post frequency by hour (converted to the local timezone) per post type.}}
	\label{fig:daily-posts-per-type}
\end{figure}

In addition to ad disclosure strategies, we also analyse post-timing patterns to better understand how sponsored content differs from regular posts. 
Content creators might strategically select the time of day they post to maximise engagement, especially for sponsored content. 
\autoref{fig:daily-posts-per-type} shows the post-frequency aggregate by hour of the day converted to the local timezone of the corresponding country; for countries with multiple timezones, we select the one that covers the most population. 
We compare sponsored and non-sponsored posts to verify whether influencers employ different strategies according to the type of content.

The trend for both types of posts is consistent during most of the day; however, there is a significant increase of sponsored posts in the late afternoon (between 5 pm and 9 pm), which coincides with prime time for legacy media advertising hours. 
This result confirms that there is a clear pattern for when content creators post most sponsored content, indicating that optimising post-timing is, in fact, (still) a marketing strategy. 
Understanding this strategy could be useful for law enforcement: focusing on analysing activity around this time window could lead to identifying more disclosed and undisclosed sponsored content.

\section{What is Monetised?}
\label{sec:brands}

We analyse the brands that influencers tag in their sponsored posts to understand \textit{what} they advertise and gain more insight into their marketing strategies. We use the \textit{paid partnership with} label, which requires influencers to include the sponsoring brand's username, to obtain a reliable list of brands by compiling all these usernames. 

In total, 1752 unique brand accounts were tagged in posts in our dataset using the paid partnership label; 1490 (85\%) of them are also explicitly mentioned in the caption. 
This number comprises 33.4\% of all users tagged in the disclosed posts. 39.4\% of all disclosed sponsored posts tag at least one of these brands, indicating that we compiled a representative sample. The ten most common brands found in our dataset are Embelleze, Eudora, Gymshark, Adidas, Zalando, Natura, PrettyLittleThing, LBA Sunglasses Boutique (@lbdashop), Intimissi, and Huawei. These brands span a variety of sectors, including fashion, beauty, sports, and electronics. To delve deeper into which brands collaborate with micro and mega influencers, \autoref{tab:brands-size} presents the most frequently tagged brands for each influencer category.

\begin{table}[htbp]
\caption{Usernames of the ten most frequently tagged brands per influencer category.}
\centering
\begin{tabular}{@{}cc@{}}
\toprule
\textbf{Micro}           & \textbf{Mega}  \\ \midrule
embelleze                & eudora         \\
zalando                  & gymshark       \\
prettylittlething        & adidas\_de     \\
intimissimibrasiloficial & embelleze      \\
naturabroficial          & lbashop        \\
vivaraonline             & huaweimobilepl \\
forage                   & cea\_brasil    \\
picsart                  & riachuelo      \\
fojisportsofficial       & womensbest     \\
samsungbrasil            & aboutyoude     \\ \bottomrule
\end{tabular}
\label{tab:brands-size}
\end{table}

Among the top ten brands, only Embelleze  -- a Brazilian cosmetics brand -- appears in both micro and mega influencer lists. Taking into account the entire dataset, there is an overlap of 197 brands between the two categories, representing 11.24\% of the overall distribution. This overlap highlights a substantial difference in most brand partnerships across influencer categories. This variation emphasises the unique brand strategies for each influencer group.

We observe significant differences in the distribution of brands across countries. The posts of Brazilian influencers tag the highest number of brands: 688 (46.6\% of the total of our compiled list), while US, DE, and NL, respectively, tag 511 (34.3\%), 433 (29.1\%), and 207 (13.9\%); \textit{mega} influencers tag more brands than \textit{micro} in all countries. The higher number of brands tagged by Brazilian content creators indicates more activity and diversity in the context of content monetisation, as more brands opt to partner with influencers for advertising. \autoref{tab:brands-size-country} shows the ten most frequently tagged brands by influencers from each country in our dataset. 

\begin{table}[htbp]
\caption{Usernames of the ten most frequently tagged brands per country.}
\centering
\tiny
\begin{tabular}{@{}cccc@{}}
\toprule
\textbf{US}            & \textbf{DE}    & \textbf{BR}              & \textbf{NL}              \\ \midrule
prettylittlething      & adidas\_de     & embelleze                & gymshark                 \\
americanexpress        & zalando        & eudora                   & womensbest               \\
bangenergy             & huaweimobilepl & naturabroficial          & mavenbeauty              \\
fijiwater              & aboutyoude     & lbashop                  & nakdfashion              \\
sinfulcolors\_official & forage         & intimissimibrasiloficial & fojisportsofficial       \\
nordstrom              & adidasfootball & cea\_brasil              & zalando                  \\
tmobile                & 4f\_official   & riachuelo                & natur\_e\_indonesia      \\
adobestock             & lorealparis    & vivaraonline             & millionairequeenswimwear \\
lyft                   & picsart        & samsungbrasil            & fellegance               \\
netaporter             & gymshark       & oboticario               & milka\_netherlands       \\ \bottomrule
\end{tabular}
\label{tab:brands-size-country}
\end{table}


To better understand which brands (e.g., local or international, type of industry) influencers partner with, we measure the general prevalence of brands across different countries. 241 brands (16.2\%) occur in posts by influencers from two different countries; 87 (5.8\%), from three; and 21 (1.4\%) are tagged in all countries. 
Therefore, at least 23.4\% of brands tagged in our dataset are international, and posts tagging these brands comprise 17.6\% of the total volume of disclosed posts. These brands are predominantly from the fashion and beauty industries (e.g. Gucci, Dior, Zara, Lancôme). 
We analyse the distribution of these posts across \textit{mega} and \textit{micro} influencers to understand brand preference when selecting the influencers to collaborate with: 66.1\% are by \textit{mega} influencers and 33.9\%, by \textit{micro}, suggesting that international brands may prefer to partner with influencers with larger audiences. 
We also analyse the distribution by setting (country + size) and observe that \textit{mega} influencers have more posts sponsored by global brands in all countries, with the highest prevalence in Germany; however, German \textit{micro} influencers still have proportionally more posts than all other settings. 
This result reinforces the evidence for the presence of undisclosed sponsored posts from other countries, as the real distribution of disclosures should be closer to the German one because of its stricter regulations.

To further investigate the potential presence of undisclosed sponsored posts, we analyse the overall prevalence of tagged brands in the entire dataset, including posts without disclosures. 
In total, 51.6K posts have at least one of the brands tagged; only 13.1K of them (25.4\%) are disclosed as sponsored.
Therefore, 74.6\% of these posts could be undisclosed ads. Moreover, 362 out of the 400 accounts in our dataset (90.5\%) have at least one post tagging one of these 1490 brands, which shows that brands engaging in sponsored content have massive reach across all settings. 
We also analyse the distribution of undisclosed posts tagged with brands considering country and size: the majority (58.5\%) are posted by BR influencers (31.4\% \textit{mega} and 24.6\% \textit{micro}), followed by US \textit{mega} (20.8\%), NL \textit{mega} (10.9\%), US \textit{micro} (4.4\%), NL \textit{micro} (3.7\%), DE \textit{mega} (2.8\%), and DE \textit{micro} (1.4\%). 
This result indicates that the stricter disclosure regulations for German influencers are effective, as most posts tagged with these brands are actually disclosed as ads. On the other hand, Brazilian content creators disclose the least. Although a post could mention a brand without necessarily being sponsored by it, our dataset likely contains a significant volume of undisclosed ads. 

\section{Impact of Disclosures on Engagement}
\label{sec:engagement}

Research highlights that influencers often hesitate to disclose sponsored posts adequately, fearing that they will get less engagement, thus decreasing their value~\cite{christinDramaMetricsStatus2021,cotterPlayingVisibilityGame2019,arriagadaYouNeedLeast2020}. Ershov and Mitchell~\cite{ershovEffectsInfluencerAdvertising2020} concluded that after the strengthening of disclosure regulations in Germany engagement with disclosed posts fell substantially. Engagement metrics often determine an influencer's value for a brand; therefore, reductions in engagement are likely to impact the monetisation potential of content creators.

We define engagement as the number of likes and comments averaged over all posts of the posting influencer. We do not normalise engagement by the number of followers because this value is only available when data are retrieved; thus, we do not have access to the number of followers an account had when each post was published. We conducted experiments to verify whether stricter disclosure requirements reduce engagement and to understand which influencers these reductions affect the most. We focus on two types of sponsored posts: disclosed (\autoref{sec:engagement-disc}) and undisclosed (\autoref{sec:engagement-undisc}).

\subsection{Engagement with Disclosed Sponsored Posts}
\label{sec:engagement-disc}
\begin{figure}[htbp]
	\centering
	\includegraphics[scale=0.55]{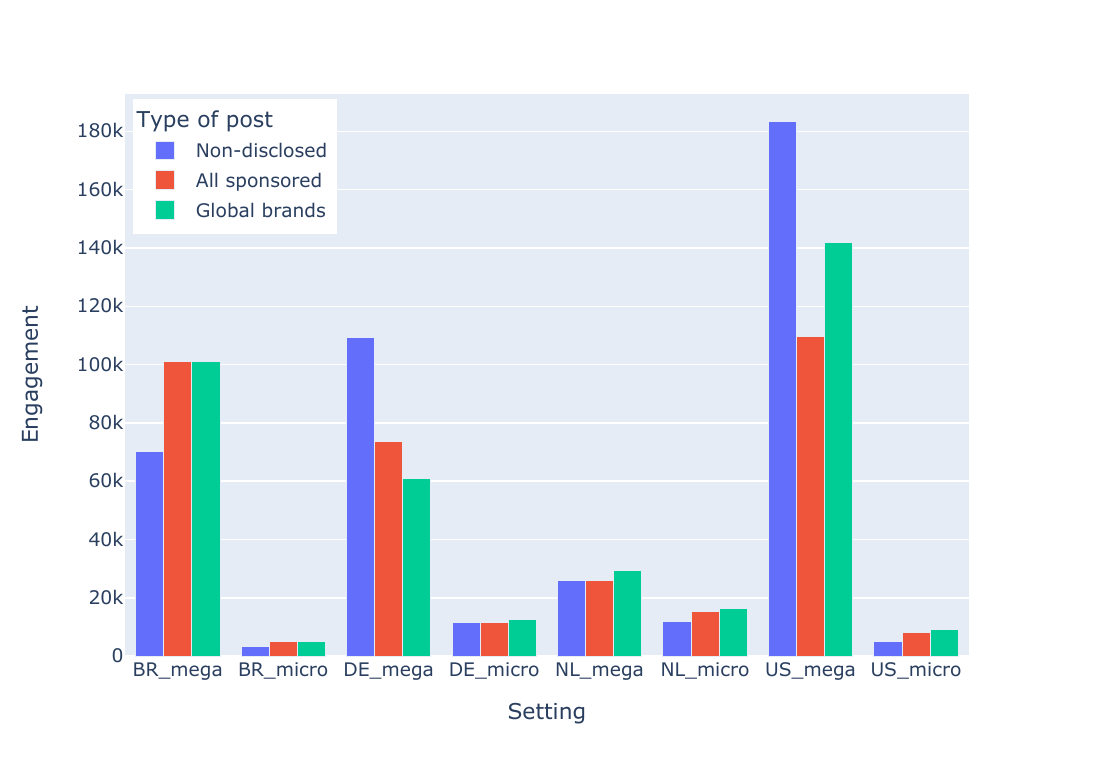}
	\caption{{Average engagement per post across different settings.} Posts are split into disclosed sponsored posts, non-disclosed posts, and sponsored posts tagging global brands. Global brands are identified from the ``paid partnership with'' feature and are present in all countries.} 
	\label{fig:engagement-per-setting}
\end{figure}

We start by comparing the average engagement in three scenarios: non-disclosed posts, all disclosed posts, and disclosed posts tagging global brands -- as these brands are widely recognised, we investigate whether posts tagging them have more engagement than sponsored posts with local brands. \autoref{fig:engagement-per-setting} shows the average engagement value per setting, highlighting significant differences across countries. 

For mega-influencers from DE and US, engagement with non-disclosed posts is higher, while sponsored posts have a higher engagement for Brazilian accounts. The differences between NL accounts and micro-influencers are generally smaller, and engagement tends to be consistent regardless of disclosures. Moreover, posts tagged with global brands generally have more engagement. 

\begin{figure}[htbp]
	\centering
	\includegraphics[scale=0.55]{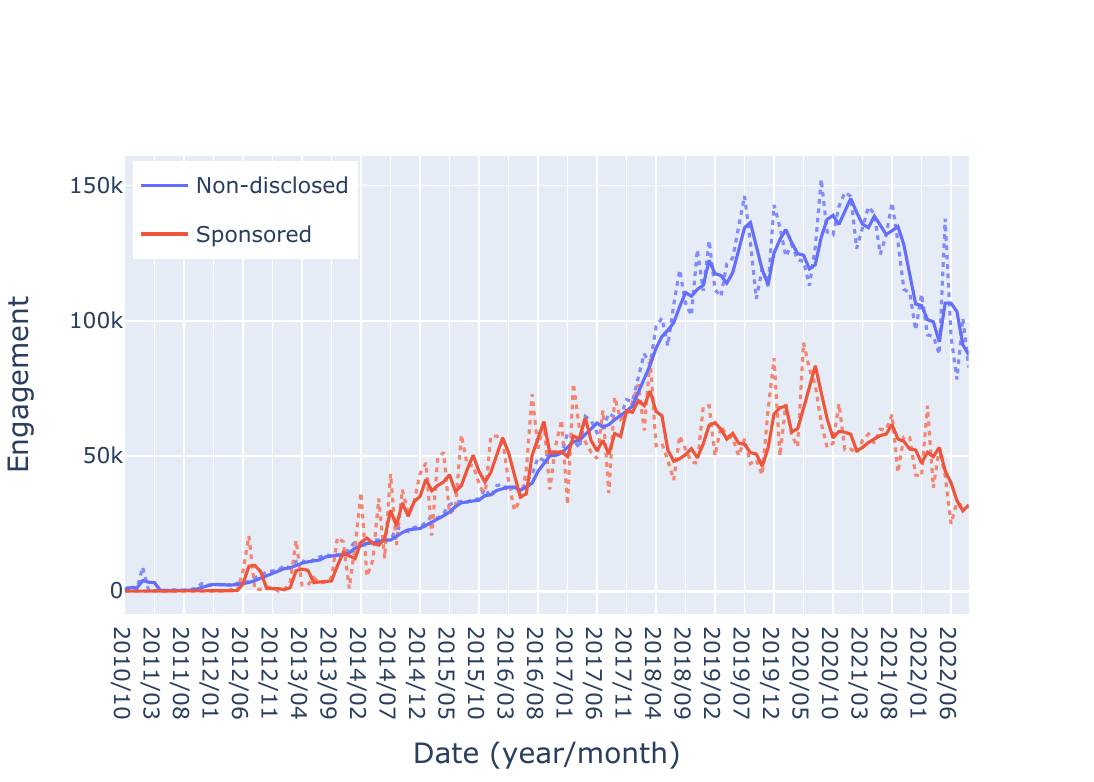}
	\caption{{Engagement trend with non-disclosed and sponsored posts over time across the entire dataset.} The solid lines represent the smoothed average engagement over a time window of 3 months. The dotted lines represent the monthly averages.}
	\label{fig:engagement-general}
\end{figure}
\begin{figure}[htbp]
	\centering
	\includegraphics[scale=0.55]{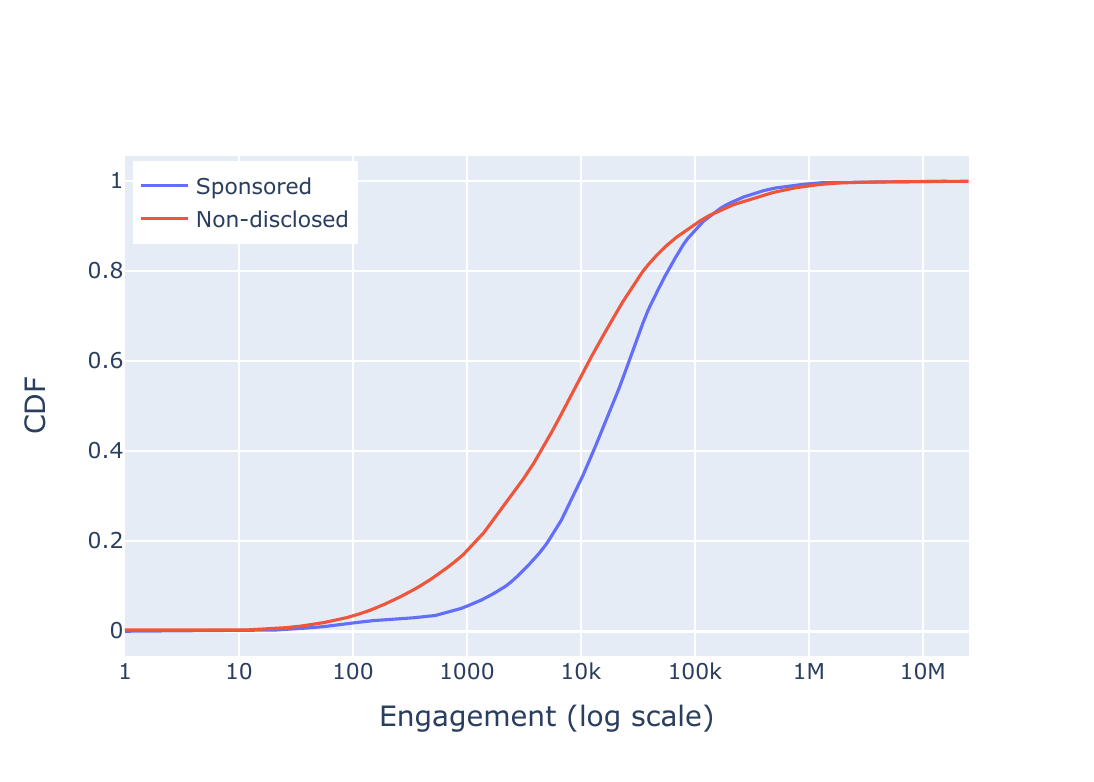}
	\caption{{Cumulative distribution of engagement with non-disclosed and sponsored posts across the entire dataset over time. The x-axis uses a logarithmic scale for clarity.}}
	\label{fig:engagement-cdf-general}
\end{figure}

To analyse fluctuations in engagement through time, we calculate the average monthly engagement with (disclosed) sponsored and non-disclosed posts. First, we analyse the general engagement trend in the dataset by aggregating data from all countries. \autoref{fig:engagement-general} shows the results; \autoref{fig:engagement-cdf-general} presents a Cumulative Distribution Function (CDF) plot to describe engagement from a more granular perspective. 

The engagement trends show that non-disclosed posts gain engagement more rapidly than sponsored ones, with initial higher engagement for sponsored posts shifting over time due to occasional spikes rather than consistent levels. The CDF plot reveals faster growth in engagement for non-disclosed posts; for example, at 1,000 and 10,000 engagement levels, non-disclosed posts outpace sponsored posts in reaching moderate engagement levels -- both types of posts plateau in engagement beyond the 100k mark, indicating a ceiling effect. Analysis of monthly cumulative averages pinpoints July 2018 as the turning point where non-disclosed posts consistently outperformed sponsored posts in engagement.

To understand the reasons behind the spikes in engagement with sponsored posts, we identified months where sponsored posts outperformed non-disclosed ones in average monthly engagement. Our goal was to determine whether specific accounts with high levels of engagement were driving these spikes. To achieve this, we identified accounts considered outliers in those months using the Interquartile Range (IQR) method. We aggregated the total engagement for sponsored posts and then calculated each account's monthly engagement and their proportional contribution to overall engagement. Finally, we applied the IQR method, identifying outliers by computing the difference between a distribution's $75^{th}$ and $25^{th}$ percentiles and flagging any account that deviates more than $1.5*IQR$ from this range as an outlier.

We found that 88.9\% of the months with spikes in engagement had outliers, with an average of 5.3 outlier accounts per month. In comparison, 82.0\% of the months in the dataset had outliers, with an average of 12.8. The difference in the average number of outliers per month is significant and shows that months with spikes are much more concentrated and dominated by a few accounts. 89.7\% of the outlier accounts were mega-influencers, confirming that content creators with more followers generate proportionally more engagement and that it is more difficult for micro-influencers to obtain above-average engagement with sponsored posts.


Our analysis initially focused on monthly outliers but missed consistent high performers. To address this issue, we applied the IQR method on an account level. We calculated the engagement contribution percentage per account for all months and applied the IQR threshold to identify outlier months. We defined a given account as \textit{outperforming} in a month if it was considered an outlier using both methods: their contribution in a specific month was an outlier, and that specific month was also an outlier. We identified 27 accounts consistently outperforming in engagement spikes, with 69.6\% being mega-influencers. This result suggests that mega-influencers more frequently achieve and sustain high engagement compared to micro-influencers. Our outlier detection accounts for individual influencer trends, ensuring a balanced comparison across different audience sizes.

To understand how engagement is distributed across the accounts in the dataset, we analysed the posts within the last spike (March 2018), which had only one overperforming account: Anitta, a Brazilian singer and a mega-influencer in our dataset. She had nine sponsored posts that month and contributed 12.7\% to the total engagement. Interestingly, 8 out of 9 sponsored posts tagged the same brand (O Boticário) a Brazilian cosmetics brand. All these posts were advertising the release of a new music video sponsored by the brand. Therefore, the spike in engagement can be explained by a specific event (the music video release) and a marketing campaign (the partnership with the brand). This result indicates that spikes in engagement can help detect marketing campaigns or related events, which might be relevant for regulating and enforcing disclosures. Furthermore, Duffy et al.~\cite{duffy2021value} highlight that influencers use strategies to actively encourage followers to interact, for example, by asking users to comment to participate in \textit{giveaways}; such strategies have been shown to inflate, at least temporarily, engagement~\cite{cotterPlayingVisibilityGame2019}. 

Our analyses generally confirm the decrease in average engagement with sponsored posts over time observed in related work~\cite{ershovEffectsInfluencerAdvertising2020}. However, this decrease (which, in our dataset, significantly sharpens from July 2018) only concerns engagement with non-disclosed posts. The spikes in average monthly engagement in sponsored posts are mostly driven by mega-influencers who consistently have high engagement or other popular accounts with overperforming monthly engagement. These results indicate that mega-influencer followers are more stable and tend to engage more consistently regardless of disclosures. However, achieving above-average engagement with sponsored posts is more difficult for micro-influencers. Therefore, they are more likely to be impacted by changes in disclosure regulation and can suffer more negative consequences from the loss of engagement because they are less likely to have overperforming posts. Thus, it is essential to analyse the implications of disclosure regulation for these categories separately. 

\begin{figure}[htbp]
	\centering
	\includegraphics[scale=.45]{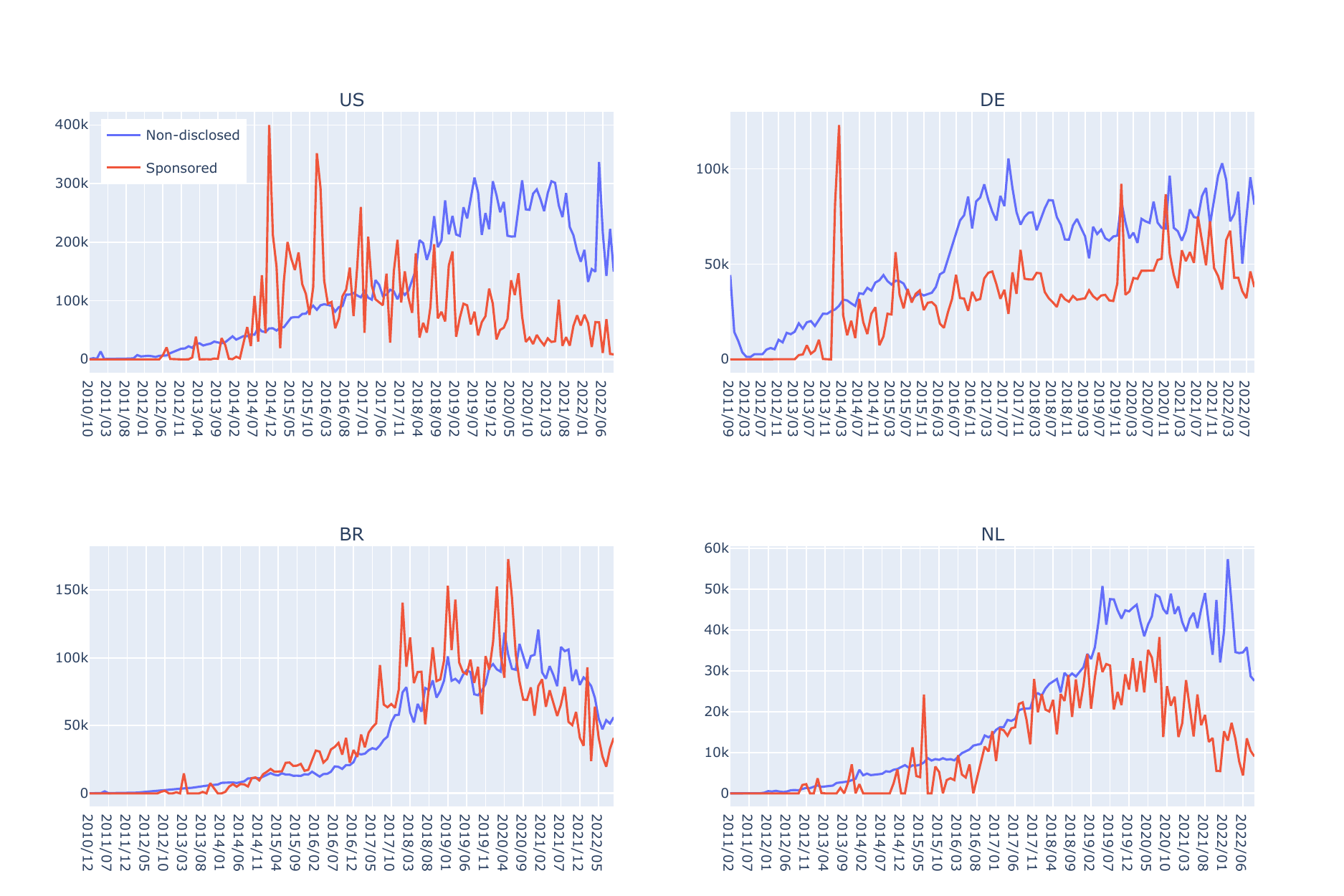}
	\caption{{Engagement with non-disclosed and sponsored posts over time per country.}}
	\label{fig:engagement-per-country}
\end{figure}

We analyse the average monthly engagement per country to investigate whether these observations are consistent across countries. \autoref{fig:engagement-per-country} shows the results. 

For US, DE, and NL, the engagement trend is similar to the entire dataset: engagement with sponsored posts starts higher with many spikes but stops growing at some point. For BR, engagement with sponsored posts remains closer to non-disclosed ones, although there is an increase in the difference in the later months. To measure the point at which the trend changes, we first calculated the monthly difference between non-disclosed and sponsored posts. Then, we applied a rolling average with a window of 3 and found the last month in which the average was negative (i.e., higher for sponsored posts) than the previous rolling average (corresponding to the previous three months). We define this month as the inflexion point in the trend. The inflection points are October 2015 for DE, January 2018 for NL, Jan 2018 for US, and August 2020 for BR.

The averages before and after the inflexion points are different in scale but comparable. Engagement with sponsored posts was higher before the inflexion point but significantly dropped and was surpassed by non-disclosed posts. The spikes in average monthly engagement still occur after the inflexion point, mainly caused by engagement with posts by mega-influencers. The specific dates are connected to the strictness of disclosure regulations in each country: they come much earlier in Germany, later and at the same time for the US and Netherlands, and only recently for Brazil. The dates are defined based on a rolling average with a window of 3, so they are still susceptible to local and short-range variations, potentially reflecting local events or seasonal factors. For future work, we plan to explore more robust methods for measuring the changes in trends and further investigate the causes for each inflexion point.

\subsection{Engagement with Undisclosed Sponsored Posts}
\label{sec:engagement-undisc}
Our previous analyses focused on self-disclosed posts. However, it is common for posts without disclosures to actually be undisclosed advertisements. Previous research has shown that the prevalence of undisclosed ads on Instagram is significant; Zarei et al.~\cite{zareiCharacterisingDetectingSponsored2020} estimate that 17.7\% of all posts contain undisclosed ads. This result raises the concern that undisclosed advertisements could influence the engagement trends observed. To better understand their impact, we used machine learning models to automatically detect undisclosed sponsored posts and assess their impact on engagement trends.

\subsubsection{Detecting Undisclosed Sponsored Posts}
We frame the detection of undisclosed ads as a semi-supervised classification problem. Given a post without any disclosures, our objective is to determine whether it is a sponsored advertisement or a non-sponsored post. To approach this problem, we use models trained on a dataset built explicitly for this task. For our classification task, we use disclosed sponsored posts, including all types of disclosures, as positive examples for the \textit{ad} (sponsored post) class. The remaining posts are considered negative examples: either \textit{non-sponsored} or \textit{undisclosed-ads}. We do not have explicit labels for all classes in the semi-supervised setting -- in practice, the models learn to detect ads based on disclosed sponsored posts. To evaluate whether models can also identify undisclosed ads, we manually label a sample of the dataset containing undisclosed sponsored posts; we will describe the annotation process and the labelled sample in~\autoref{sec:gold-data}. 

Our original dataset contains 34.7K disclosed posts and 971.5K negative examples, which results in a highly imbalanced distribution of classes. We apply the random undersampling approach proposed by~\cite{zareiCharacterisingDetectingSponsored2020} to address this imbalance and create a balanced dataset for our classification task. This approach involves selecting all disclosed posts and a random subset of the same size from the negative class. This results in a balanced dataset with an equal 50/50 split between the two classes. This balanced distribution differs significantly from the heavily skewed actual distribution; however, we draw the annotated sample used for evaluation from the real distribution, so evaluating the model on it allows us to investigate how it would perform on real-world data. Finally, we create separate datasets for each country (BR, DE, NL, US) with 20.6k, 35.1k, 3.5k, and 11.2k posts for each and a combined dataset with 70.4k posts. We evaluate the models in three steps: first, we use 80\% of the dataset for training and 20\% for evaluation; then, we perform 5-fold cross-validation to ensure that the random split between the train and test sets does not affect the model performance. Finally, we evaluate the models on the annotated data sample, which includes a representative group of posts, including undisclosed ads.

Our models use a post's caption (text) as the input signal. We remove all disclosure hashtags and keywords from the captions to ensure that the models are not learning a direct mapping between disclosures and advertisements. We do not pre-process the text further. We used a BERT-based model for classification~\cite{devlin-etal-2019-bert}. Using our dataset, we build our classifier using a pre-trained BERT model that is fine-tuned for the ad detection task. We used the \textit{bert-base-multilingual-uncased}\footnote{\url{https://huggingface.co/bert-base-multilingual-uncased}} pre-trained model, which is trained on text from multiple languages and can handle the mix of languages in our dataset. We used the pre-trained model weights from HuggingFace and their library to implement all of our experiments~\cite{wolf-etal-2020-transformers}.

We fine-tuned the BERT-based model using the same hyperparameters from the original paper (specified in~\cite{devlin-etal-2019-bert}) and trained it for three epochs. To evaluate the impact of country-specific features, we train models using two settings: one model per country and one model using the combined dataset. We hypothesise that using the combined dataset will allow the model to learn common patterns across the posts by influencers from different countries in the dataset, resulting in a more robust and generalisable model. We evaluated our models using precision, recall, and F1. \autoref{tab:clf-single-eval} shows the performance of the models trained on individual country datasets using cross-validation. All metrics are the average across the five folds; values after $\pm$ indicate the standard deviation across folders.

\begin{table}[htbp]
\caption{Performance of the models trained on each country's individual dataset with 5-fold cross-validation, evaluated using Precision, Recall, and F1. \# Samples represents the number of posts in the test set of each country. All metrics are averaged over the five folds; values after $\pm$ indicate standard deviation.}
\centering
\begin{tabular}{@{}ccccc@{}}
\toprule
\textbf{Dataset} & \textbf{P}                      & \textbf{R}                      & \textbf{F1}                     & \textbf{\# Samples} \\ \midrule
\textbf{BR}      & .89 $\pm$ .01 & .90 $\pm$ .03 & .90 $\pm$ .02 & 4130                \\
\textbf{DE}      & .93 $\pm$ .03 & .84 $\pm$ .07 & .88 $\pm$ .04 & 7006                \\
\textbf{NL}      & .88 $\pm$ .04 & .90 $\pm$ .03 & .89 $\pm$ .02 & 693                 \\
\textbf{US}      & .86 $\pm$ .03 & .89 $\pm$ .03 & .87 $\pm$ .03 & 2247                  \\                
\bottomrule
\end{tabular}
\label{tab:clf-single-eval}
\end{table}

Classifiers trained in individual datasets generally perform well, with Macro F1 values ranging from 0.87 to 0.90 and an average of 0.885. The performance across countries varies, with the best result seen in the model trained on the German (DE) dataset. The higher number of disclosures in Germany results in a larger, more representative training dataset, contributing to this result. The F1 values show that the models can effectively classify sponsored posts and negative examples without a strong bias toward one class. However, there are differences between the precision and the recall values. The models for DE tend to be conservative in identifying sponsored posts, with higher precision but lower recall. In NL and US, the opposite occurs, leading to an overestimation of the prevalence of ads and misclassification of non-sponsored posts. Despite these differences, the F1 scores are balanced, and performance is consistent across classes. 

We analyse the models' performance on the combined dataset using the same metrics. We first evaluate the model using the test split (with 20\% of the data); the combined dataset is much larger than the individual ones; thus, training the model on it is more costly and performing cross-validation is inefficient. Therefore, to save computational resources, we use the 80/20 split to train the model and more quickly check whether the use of the combined data improves the classification performance. \autoref{tab:clf-combined-eval} presents the results.

\begin{table}[htbp]
\caption{Performance of the models trained on the combined dataset evaluated on the test set (20\% of the dataset used in the classification task).}
\centering
\begin{tabular}{@{}ccccc@{}}
\toprule
\textbf{Dataset} & \textbf{P}                      & \textbf{R}                      & \textbf{F1}                     & \textbf{\# Samples} \\ \midrule
\textbf{BR}      & .85 & .86 & .85 & 4130                \\
\textbf{DE}      & .89 & .89 & .89 & 7006                \\
\textbf{NL}      & .84 & .83 & .83 & 693                 \\
\textbf{US}      & .80 & .80 & .80 & 2247                  \\                
\bottomrule
\end{tabular}
\label{tab:clf-combined-eval}
\end{table}

The model trained on the combined dataset shows a decline in performance in all countries except Brazil, for which the model achieves a marginal increase of only one point in F1. The differences in disclosure practices and marketing strategies among countries make it challenging for the model to generalise effectively. Moreover, training the model on the combined dataset takes longer and requires more computational resources. Therefore, we used the models trained on individual country datasets for all subsequent experiments and chose not to evaluate the combined model using cross-validation. 

\subsubsection{Evaluation on Annotated Data}
\label{sec:gold-data}
To verify whether trained models can identify undisclosed ads, we evaluate their performance on a manually annotated sample. Annotators labelled \textit{posts} as sponsored or \textit{non-sponsored} based on their captions. We selected 1210 posts by US influencers to be labelled. For future work, we will label posts by influencers from all countries in the dataset. We randomly sampled 50\% of the total from disclosed posts and 50\% from undisclosed. A group of fifteen students in a Master's programme in Law \& Technology labelled the dataset. After a training session, all participants labelled a sample of 100 posts used to calculate the inter-annotator agreement. We measure agreement using absolute and majority agreement percentages and Krippendorf's Alpha~\cite{krippendorff2011computing}. We achieved 58\% absolute agreement (i.e., all annotators selected the same label for 58\% of the posts) and 0.74 Krippendorf's Alpha, representing good agreement. The agreement metrics indicate that the labelled data are reliable.

For the main annotation phase, we split the students into five groups of three people. Members of the same group labelled the same posts, ensuring that three people annotated each post. We used majority voting to determine the final label of a post. 36.8\% of the undisclosed posts were labelled as sponsored. In total, the annotated sample contains 224 undisclosed ads. Finally, we evaluate the performance of the ad detection model on annotated data. The model trained on US data achieves 0.80 precision, 0.80 recall, and 0.80 F1, representing a seven-point decrease in performance compared to cross-validation evaluation. The model correctly detected 151 of the 224 undisclosed ads (67.4\%). Thus, even with the drop in F1, the model learnt to identify undisclosed ads and can detect most of them. Although we only evaluate the classifier trained on US data, the ones for the other countries are based on the same pre-trained multilingual BERT model (M-BERT) and fine-tuned on comparable data (relying on self-disclosure). Therefore, their performance in undisclosed ads would be reliable and follow the same pattern. Pires et al.~\cite{pires-etal-2019-multilingual} demonstrated that M-BERT effectively processes multiple languages despite being trained without explicit cross-lingual objectives, highlighting its utility in tasks in languages such as Spanish, Hindi, and Russian. This suggests that the linguistic features learnt by M-BERT can be generalised across languages, making it a robust foundation for our classifiers in other countries. Moreover, studies have shown that M-BERT performs strongly in cross-lingual zero shot tasks~\cite{wu-dredze-2019-beto,karthikeyan2020crosslingual,devlin-etal-2019-bert}.

\subsubsection{Is Not Disclosing Worth It?}
After training the models, we use them to predict whether the posts without disclosures are ads. We apply the models to all undisclosed posts in the dataset. The model classified 123.6k posts as undisclosed ads (12.3\% of the entire dataset). \autoref{tab:spons-distribution} presents the distribution of sponsored posts across countries.

\begin{table}[htbp]
\caption{Distribution of sponsored (disclosed and undisclosed) posts per country across the dataset.}
\centering
\begin{tabular}{@{}ccccccc@{}}
\toprule
\multicolumn{1}{l}{Country} & \multicolumn{2}{c}{\textbf{Disclosed}} & \multicolumn{2}{c}{\textbf{Undisclosed}} & \multicolumn{2}{c}{\textbf{Total}} \\ \midrule
\multicolumn{1}{l}{}        & \#                 & \%                & \#                   & \%                & \#                & \%             \\
\textbf{BR}                 & 10.3k              & 2.2               & 57.0k                & 12.3              & 67.3k             & 14.6           \\
\textbf{DE}                 & 17.2k              & 16.4              & 4.8k                 & 4.6               & 220.0k            & 21.0           \\
\textbf{NL}                 & 1.7k               & 1.2               & 22.3k                & 15.4              & 24.0k             & 16.7           \\
\textbf{US}                 & 5.6k               & 1.9               & 39.5k                & 13.4              & 45.1k             & 15.3           \\ \midrule
\textbf{All}                & 34.8k              & 3.5               & 123.6k               & 12.3              & 158.3k            & 15.8           \\ \bottomrule
\end{tabular}
\label{tab:spons-distribution}
\end{table}

The results show a clear trend where most sponsored posts in the dataset lack proper disclosures, except for posts made by German influencers. The rates of undisclosed ads are uniform across countries except DE, indicating that the model predictions are consistent. The ratio between disclosed and undisclosed ads varies proportionately -- a higher number of disclosed ads corresponds to a lower number of undisclosed ones. The results are particularly striking for German influencers, where most sponsored posts are disclosed, highlighting the effectiveness of strict regulations. Additionally, the overall proportion of sponsored posts across countries demonstrates a comparable distribution. Finally, we extend the engagement analyses from \autoref{sec:engagement-disc} to include the three categories of sponsorship -- non-sponsored, disclosed ad, and undisclosed ad. First, we compare the average engagement per country for the entire dataset. 

\begin{figure}[htbp]
	\centering
	\includegraphics[scale=0.55]{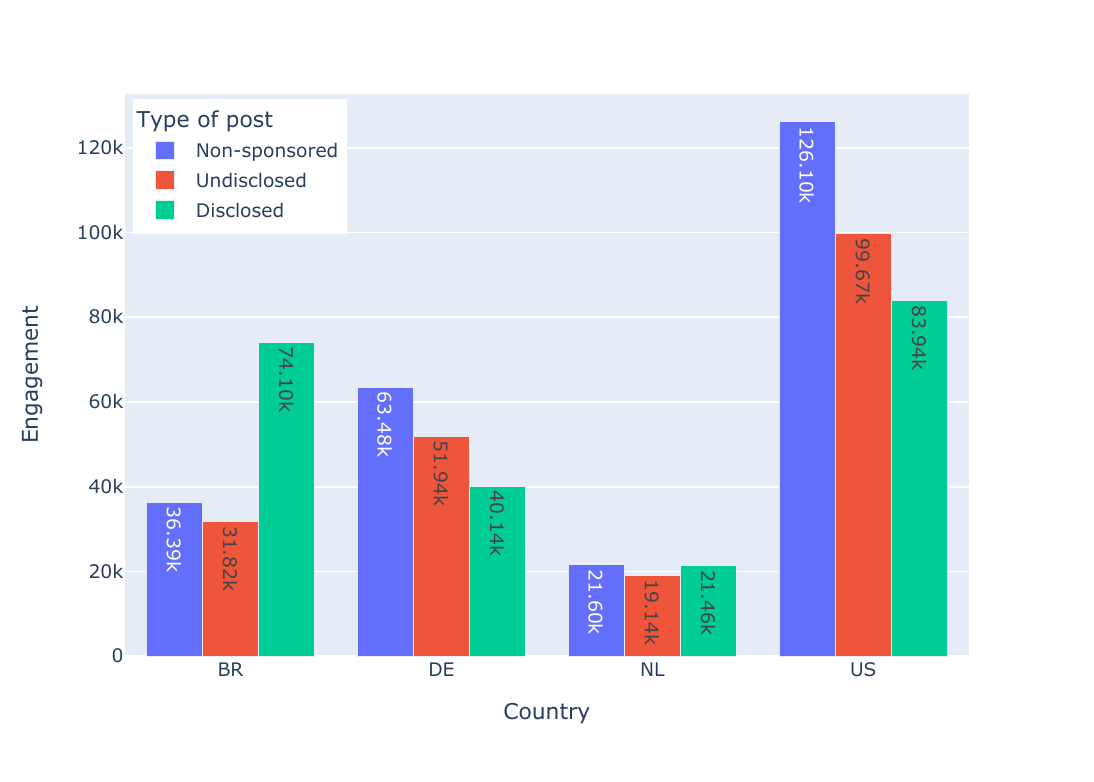}
	\caption{{Average engagement with non-sponsored, disclosed, and undisclosed sponsored posts across the entire dataset.}}
	\label{fig:engagement-per-spons-type}
\end{figure}

\begin{figure}[htbp]
	\centering
	\includegraphics[scale=0.55]{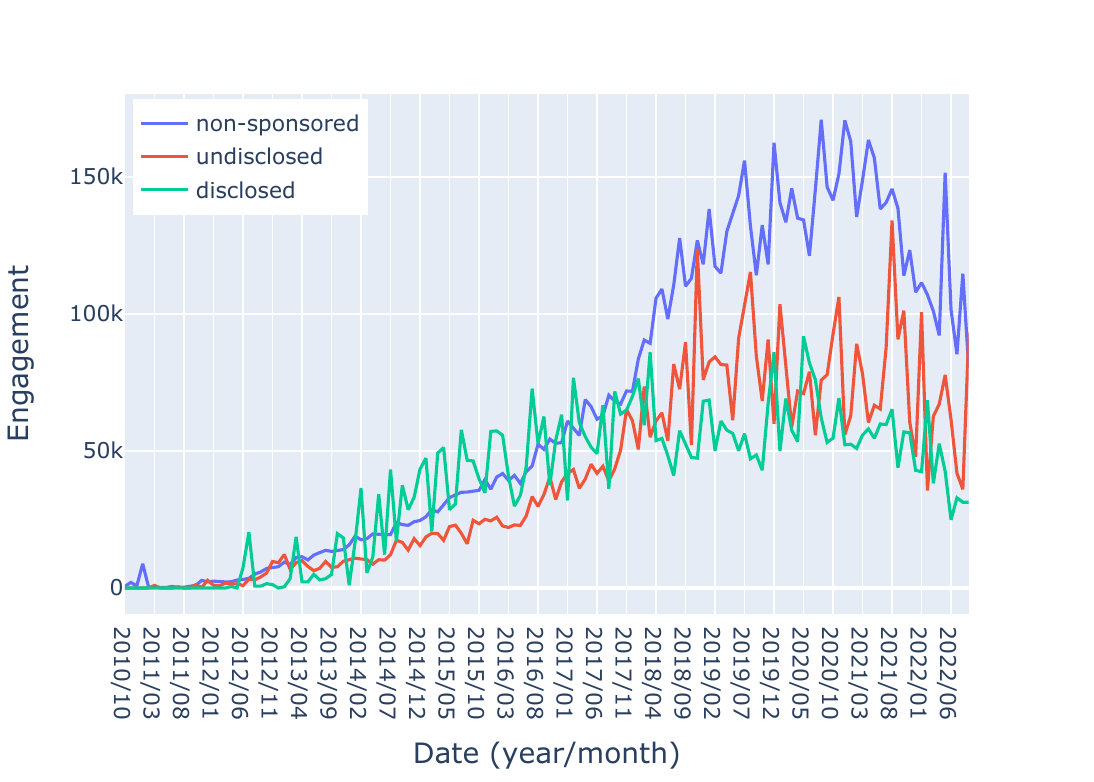}
	\caption{{Average engagement with non-sponsored, disclosed, and undisclosed sponsored posts over time across the entire dataset.}}
	\label{fig:engagement-spons-type}
\end{figure}

\autoref{fig:engagement-per-spons-type} shows that the average engagement varies significantly across countries. Non-sponsored posts receive more engagement, except for posts by Brazilian influencers. In Brazil, engagement is higher for disclosed sponsored posts than undisclosed ones, whereas the reverse is true for Germany and the US. The difference in the Netherlands is negligible. Disclosed sponsored posts receive an average engagement of 56.3k when aggregating the results across all countries. In contrast, undisclosed ads receive 52.0k, suggesting that proper disclosure does not necessarily result in a significant decrease in engagement. The decrease in engagement is more evident between sponsored and non-sponsored posts, indicating that the presence of ads results in lower engagement rather than the disclosure itself. To describe how engagement trends change through time, \autoref{fig:engagement-spons-type} presents the time series of average engagement across the entire dataset.

The engagement trends follow the pattern described by~\autoref{fig:engagement-general}. Engagement with non-sponsored posts grows proportionally much more than with the others. We previously hypothesised that undisclosed ads could inflate engagement with non-disclosed posts, leading to their apparent higher growth. However, \autoref{fig:engagement-spons-type} confirms that the trend remains consistent when splitting non-disclosed posts into non-sponsored and undisclosed ads. Although engagement with undisclosed ads has many spikes, bringing the average up, generally the values are lower than the ones for non-sponsored posts and more similar to disclosed ads. These spikes follow the behaviour we discussed previously -- accounts with many followers drive engagement up, especially through marketing campaigns. Overall, these trends confirm our observations that sponsored posts have lower engagement. Moreover, except for spikes that cause the average to increase, engagement with disclosed and undisclosed ads remains close, indicating that properly disclosing ads is not necessarily harmful to engagement.  




\section{Summary}

This paper presents a longitudinal characterisation of self-disclosed sponsored content posted on Instagram by 400 influencers from four countries from the platform's release in October 2010 to September 2022. Our study revealed the following observations: First, our longitudinal analysis highlights the impact of adopting new features of platform monetisation in different countries. 
Specifically, we observe that the introduction of the \emph{paid partnership with} feature in 2017 led to faster adoption in Brazil and the Netherlands and very low adoption in Germany, perhaps due to German case law and the adoption of pragmatic practices by law firms who advised influencers \enquote{to be on the safe side}~\cite{knitter_2020}. 

We also observe a significant difference in disclosure strategies across the different countries, such as the length of captions for disclosed posts and the position of the disclosing term.
German case law recommends that the disclosure terms be clearly visible and separate from a hashtag cloud, which might explain the overwhelming adoption of keywords in sponsored posts by German influencers and their positioning at the beginning of the caption. Therefore, we conclude that strict disclosure requirements lead to more disclosed posts and more visible disclosures.

Second, we show that the posting activity of the observed accounts increased significantly until 2016 and then decreased steadily to the current day. The start of this decline coincides with the release of the Stories feature. We hypothesise that influencers diversified their activity by posting more Stories than permanent posts. However, the proportion of sponsored content remained constant in about 5\% of the posts in three of the countries we analysed, except for Germany, which saw a significant decrease in the percentage of disclosed posts after 2018. This result shows that influencers continued to use permanent posts to share sponsored content. 


We also observed that Brazilian accounts post significantly more than influencers from other countries, and the share of sponsored content is spread across more accounts, showing that the market is more uniform. However, high posting activity leads to lower engagement on average; we observe that the volume of posts is inversely proportional to engagement. 

These findings indicate the need for nuanced regulatory strategies that consider changes in platform characteristics and cultural differences in influencer behaviour. Additionally, these results underscore the methodological challenges in replicating and generalising findings over time, as ongoing changes in platform features and user behaviour may render past observations and trends non-representative.

Third, we identify several monetisation strategies. 
Regarding the timing of posting sponsored content, we see a significant increase in the late afternoon local time (between 5pm and 9pm), consistent with legacy media advertising hours. Therefore, the timing of posts can be a relevant feature for both machine learning and computational methods aimed at detecting sponsored content and for regulatory agencies monitoring these activities.


We thoroughly investigate the impact of sponsored content and disclosures on engagement. We identified that the engagement trend changed considerably after July 2018. Before this date, disclosed sponsored posts had higher engagement than non-disclosed posts. After July 2018, this trend reverses. We analyse the period in which the disclosed posts had higher engagement and conclude that popular mega influencers drive the average engagement up through a few posts with outlier performance; we identify coordinated marketing campaigns among these posts. This observation could be helpful for the enforcement of disclosure regulations in social media, as it suggests that monitoring spikes in posts related to specific brands or campaigns could lead to undisclosed sponsored content.

To analyse whether undisclosed ads impact the engagement trend with non-disclosed posts, we fine-tune a BERT-based model to detect sponsored content, including undisclosed ads. The model reaches a 0.90 F1 score for disclosed sponsored posts, with a 67.4\% accuracy rate when identifying undisclosed ads. We identified 123.6k undisclosed ads, representing 12.3\% of all posts. We compare the differences in engagement between disclosed and undisclosed sponsored posts and conclude that, generally, disclosures do not harm engagement. 

Our observations highlight the impact of legal uncertainty on advertising disclosure. They also reflect the importance of clear mandatory guidelines for effective disclosure strategies. Without them, even regulations that are supposed to be harmonised at the European level seem to be perceived differently and applied in Member States such as Germany and the Netherlands.
Moreover, disclosed content does not show less engagement than undisclosed content in most scenarios -- this empirical observation could be used to encourage influencers and brands to follow disclosure regulations.

We acknowledge that the 400 selected influencers may not be a representative sample of the user account population with a very large following base. However, our qualitative data curation relies on clearly defined criteria for finding and selecting influencers, allowing for more systematic and reproducible experiments. Moreover, we highlight that our study, by design, focuses on descriptive and observational analyses contextualised within specific influencer categories and countries. This approach allows us to provide nuanced perspectives on disclosure practices across different social media landscapes. However, the complexity of external factors, such as seasonality and their potential impact on influencer behaviour, presents challenges in measuring statistical significance with robustness. Consequently, while our findings offer valuable contributions to understanding influencer marketing dynamics, they must be interpreted within the framework of these methodological constraints.

\subsection*{Recommendations for Platforms and Regulatory Agencies}
Our longitudinal analysis of Instagram content provides insights that platforms and regulatory agencies can leverage for more effective oversight. A primary observation is the need for precise regulations and guidelines for advertising disclosures. Although each country has unique cultural and socio-political nuances that influence disclosure trends, a foundational set of practices could improve the rate of proper disclosures. The German scenario shows that clear guidelines result in higher compliance.

As platforms continually innovate and introduce new features, it is essential to ensure that mechanisms for ad disclosure are robust and adaptable. For instance, the advent of the Instagram Stories feature prompted influencers to diversify their content strategy. Platforms and regulators should proactively anticipate these shifts and provide tools or templates to clarify ad disclosures for the audience.

Our experiments with the BERT-based model illustrate how computational methods can effectively identify undisclosed ads. Such technologies can flag potential ads and encourage proper disclosure when integrated into platforms. Although automation can be helpful, education is fundamental for effective regulation. Both social media platforms and regulatory agencies should invest in ongoing education for the influencer community. Through workshops, webinars, and incentivising best practices, we can ensure that influencers are up-to-date with the latest guidelines and understand the importance of proper disclosures.

Understanding monetisation nuances across different influencer categories is also crucial. Given the diverse strategies between micro and mega influencers, guidelines should be tailored to address each segment's unique challenges and opportunities, ensuring no group is inadvertently disadvantaged. In conclusion, fostering transparency and trust in sponsored content requires collaboration between platforms, influencers, and regulatory agencies backed by clear guidelines, technological innovation, and continuous education.


\section*{Availability of data and materials}
All the code relevant to the paper is publically available on GitHub. We provide a script for our experiments' data collection and processing phases. This script is self-contained; researchers simply need to provide their CrowdTangle API key to initiate the data collection process in line with our methodology. The code necessary to reproduce all of our experiments is also publicly available. All resources can be found on \url{https://github.com/thalesbertaglia/instagram-disclosure-trends}. 

\section*{Funding}
This research was funded by Studio Europa, Maastricht University, and the ERC Starting Grant HUMANads.

\section*{Competing interests}
  The authors declare that they have no competing interests.

\section*{Author's contributions}
TB, CG, and JS collected and curated the dataset. TB, AI, and JS designed the experiments. TB implemented the experiments. All authors contributed to writing and reviewing the manuscript.


\bibliographystyle{plain} 
\bibliography{main}      









\end{document}